\documentclass[runningheads]{llncs}
\usepackage[T1]{fontenc}
\usepackage{graphicx}
\usepackage{amsmath}
\usepackage{caption}
\usepackage{subcaption}
\newcommand*{\affaddr}[1]{#1} % No op here. Customize it for different styles.
\newcommand*{\affmark}[1][*]{\textsuperscript{#1}}
\newcommand*{\mail}[1]{\texttt{#1}}
\begin{document}
\title{A New High-Performance Approach to Approximate Pattern-Matching for Plagiarism Detection in Blockchain-Based Non-Fungible Tokens (NFTs)}
\titlerunning{Fast [...] Plagiarism Detection in Blockchain-Driven NFTs}
\authorrunning{C. Pungila et al.}
\author{%
Ciprian Pungilă\affmark[1], Darius Galiș\affmark[1], Viorel Negru\affmark[1,2]\\
\affaddr{\affmark[1] Faculty of Mathematics and Informatics, Department of Informatics}\\
\affaddr{\affmark[2] ICAM - Environmental Advanced Research Institute}\\
\mail{\{ciprian.pungila,darius.galis,viorel.negru\}@e-uvt.ro}\\
\affaddr{West University of Timișoara, Romania}
\institute{}
}

\maketitle              % typeset the header of the contribution
\begin{abstract}
We are presenting a fast and innovative approach to performing approximate pattern-matching for plagiarism detection, using an NDFA-based approach that significantly enhances performance compared to other existing similarity measures. We outline the advantages of our approach in the context of blockchain-based non-fungible tokens (NFTs). We present, formalize, discuss and test our proposed approach in several real-world scenarios and with different similarity measures commonly used in plagiarism detection, and observe significant throughput enhancements throughout the entire spectrum of tests, with little to no compromises on the accuracy of the detection process overall. We conclude that our approach is suitable and adequate to perform approximate pattern-matching for plagiarism detection, and outline research directions for future improvements.
\end{abstract}
\section{Introduction}
Plagiarism is an important issue with respect to protecting intellectual property, a crucial centerpiece in the academic community, as well as in various business aspects. Plagiarism occurs whenever material is copied without the author's permission or approval. With the spread of the Internet, plagiarism becomes a growing concern to authors developing original work. The blockchain industry aims to resolve this problem through the use of non-fungible tokens (NFTs), a market that has grown exponentially in the past few years \cite{introduction-nft-marketcap}.

Detecting plagiarism could leverage costs significantly for businesses with respect to protecting their intellectual property, and could identify theft of ideas and methodologies commonly used in the academia. When it comes to protecting its rightful author, intellectual property is surely a form of art on its own. Blockchain-driven architectures with inherent NFT support, aim to protect digital art, in all its forms, through NFT-ready technology. This provides NFT-ready blockchain technology with the means to protect against various forms of intellectual property infringement.

This paper is organized as follows: Section \ref{RelatedWork} provides a deeper insight into the related work that lies as the ground foundation of this paper; Section \ref{Implementation} outlines the intrinsic details of our own implementation, and discusses various design choices made throughout its inception; Section \ref{ExperimentalResults} describes the testing methodology and real-world experiments that we have employed in order to test our hypothesises, and discusses the outcome of the approach as compared to other existing approaches; finally, Section \ref{Conclusions} summarizes the work we have employed, and describes our vision for future directions on it.

\section{Related work}
\label{RelatedWork}

\subsection{Plagiarism detection}
\subsubsection{Related research.}
Plagiarism detection is a serious issue of interest to the academia, as well as to businesses in general. Numerous papers and approaches have been proposed to leverage this issue, and improve existing techniques to identify potential scammers. For example, in \cite{relatedwork-scam}, the authors propose an approach based on tokenization and the computation of a similarity score, making use of natural language processing (NLP) techniques. In \cite{relatedwork-turkish}, authors focus on a hybrid approach: by combining the Jaccard and cosine distances, they recurse to machine learning (ML) and NLP, corroborated with text mining and similarity analysis. Similar to the previous approach, the authors recurse to tokenization. Saini et al \cite{relatedwork-textmining} had build a plagiarism detection using text mining methods and the cosine distance for computing similarity. Putri et al \cite{relatedwork-karprabin} propose a text-mining-based approach to plagiarism detection using a Karp-Rabin variant of pattern-matching, involving a slightly modified Jaccard similarity computation method and tokenization, to identify the number of common words in the two texts being compared. While the approach is able to identify matches in the two texts being compared, the approach itself is restricted to identifying common words. Approximate pattern-matching techniques in text mining help leverage the gap between partially modified copycats and original work, through a more accurate identification of close-enough wording to cause suspicion of plagiarism.

\subsubsection{Plagiarism detection through text mining.}
There are several classes and types of plagiarism. From a lexical perspective, text mining is a common approach to performing plagiarism detection. From a semantic perspective, as outlined before, there are approaches based on semantics, particularly NLP and ML, that bring the concept of "understanding" of the text contents into the picture, just as well. Our main focus in this paper is the approach based on text mining utilizing the tokenization of input data, in order to perform the mining process and assign a plagiarism degree between two different texts $A=a_{1}a_{2}...a_{n}$ and $B=b_{1}b_{2}...b_{n}$ over an alphabet $\sigma$.

\subsubsection{Similarity measures for approximate pattern-matching.} A few similarity measures are commonly used to perform text mining, in order to find approximate pattern-matches. Of these, the following are the approaches we will focus on in this paper:
\\

a) \textit{The Euclidean distance} \cite{relatedwork-euclidean} is a  granular approach to identifying similarities between two sequences of characters, by taking into account the numeric values of their corresponding ASCII codes, and computing a numerical similarity. It is defined, in normalized form, as:
\\

$
d(A,B) = 1-\frac{\sqrt{\sum_{i=1}^{n} (a_{i}-b_{i}})^2}{\sqrt{n} \times |\sigma|}
$
\\

b) \textit{The Hamming distance} \cite{relatedwork-hamming} measures the number of differences, position-wise, between the two texts. It is defined, in normalized form, as:
\\

$
d(A,B) = 1-\frac{\sum_{i=1}^{i=n} (d_{i}= 
  \begin{cases}
    1, & a_{i} \neq b_{i} \\
    0, & a_{i} = b_{i} \\
  \end{cases} )}{n}
$
\\

c) \textit{The Levenshtein distance} \cite{relatedwork-levenshtein} is a string metric for identifying the number of single-character edits required to transform one sequence of characters into another. Assuming costs $c$ are the various costs for the different operations possible (insertion, delete, substitution - with an implicit cost value of 1 for each of these), it is defined as:
\\

$
d_{i,j} = min
  \begin{cases}
    d_{i-1,j} + c_{delete}(b_{i}) & \\
    d_{i,j-1} + c_{insert}(a_{j}) & \\
    d_{i-1,j-1} + c_{substitute}(a_{j}, b_{i}) \times [a_{j} \neq b_{i} ] & \\
  \end{cases}
$

\subsubsection{Performance analysis.}
Not many research papers outline the performance of their approach in real-world experiments, especially when using very large datasets. With the growing demand for fast solutions in almost all fields of technology nowadays, there is no doubt that speed is becoming a major constraining factor with respect to existing implementations. Therefore, one particular emphasis in our approach lies on the acceleration of the approximate matching process overall.

\subsection{Blockchain technology}
Blockchain technology has come into focus in recent years with the growing public interest in Bitcoin \cite{relatedwork-bitcoin}. A blockchain is a ledger used to store data in a sequential manner time-wise, enforcing data immutability using security measures, of which the most common are cryptographic hashes. Data is organized into blocks, with each block having its own cryptographic hash that is dependent on the hash of the previous block. A block is usually comprised of various operations. Due to its inherent design, alterations of any part of data, in any given block, would determine inconsistency at the hash level for all consecutive blocks, ensuring therefore data non-repudiation.

\begin{figure}
\centering
\includegraphics[scale=0.43]{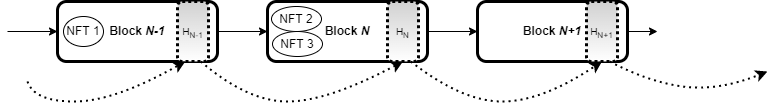}
\caption{A blockchain-driven ledger, where block $N$'s hash ($H_{N}$) is cryptographically and computationally dependent on the hash of the previous block,  $H_{N-1}$. Also depicted are various NFTs that are minted throughout blocks.}
\label{fig:blockchain}
\end{figure}

In recent years, with the growing interest in digital currencies making use of blockchain technology (such as Bitcoin), commonly known today as cryptocurrencies, the NFT market has also grown significantly in just a few months from its inception, outlining interest in use-cases of the technology for protection against copycats. Our focus in this paper is how we can apply our approach for approximate pattern-matching, to NFT technology, in particular to protecting intellectual property stored through the form of an NFT that was already previouslyminted in the blockchain.

\subsection{Challenges of NFT-driven plagiarism detection}
As blockchain technology provides nowadays the adequate means to uniquely identify and, therefore, protect intellectual property (in particular, identify plagiarized material), we have to also consider the potential drawbacks of such an approach. For example, given how hashes are able to uniquely identify a given piece of data, they are not able to differentiate between close wording, which is what approximate pattern-matching aims to achieve. With NFTs becoming more popular, a growing concern becomes the need to parse large sets of NFTs (in order to identify potential clashes or inconsistencies with other existing NFTs) whenever new NFTs are being minted. While hash techniques can offer protection against immediate copycats of data, they do not offer protection against slightly altered versions of the same data. For example, the SHA256 \cite{relatedwork-sha256} hash of the text "The sky is beautiful" differs significantly from that of the slightly altered version of the same text: "The sky's beautiful" (please refer to Table \ref{relatedwork-tab1}). Not only that, but there is no discernible link or connection between the two hashes that are computed for the two texts: by all means and purposes, different hashes mean different texts, with no indication of any semantic relationship between the texts that were hashed.

A similar principle applies to digital data in all its form (e.g. not just text), including digital art stored and valued through NFT technology. For example, a PNG picture without alpha transparency, as a form of digital art, could be digitally altered with the change of a single bit (e.g. for example, corresponding to the unused alpha transparency channel of any pixel), without any visual changes of any kind to the actual picture itself - but with completely different resulting hashes, NFT-wise, determined by the two different pieces of data. This means that the problem of uniqueness is not handled entirely in the NFT world, but could potentially be solved through an approximate pattern-matching technique, where a similarity comparison would be able to properly identify the high degree of similarity, from a lexical perspective at least, between the two files, and classify the altered file as a plagiarized version of the first one, implicitly denying the minting of the new NFT therefore.

\begin{table}
\caption{Comparison of hashes of an original text, and a slight alteration of the same original text.}\label{relatedwork-tab1}
\centering
\resizebox{\textwidth}{!}{\begin{tabular}{|l|l|l|}
\hline
\textbf{Type of text} &  \textbf{Text contents} & \textbf{SHA256 hash computed for the text}\\
\hline
Original text &  The sky is beautiful & d5217a507abdf43516559facc8b9f51cecd463fe2f4542a53ea6de3363642a62\\
Altered text &  The sky's beautiful & f452385958a7281aff97bcd2d52f9d43c29be3bcd24481f87d23e6a9b626a2a8\\
\hline
\end{tabular}}
\end{table}

\subsection{The Aho-Corasick automaton}

\begin{figure}
\centering
\includegraphics[scale=0.45]{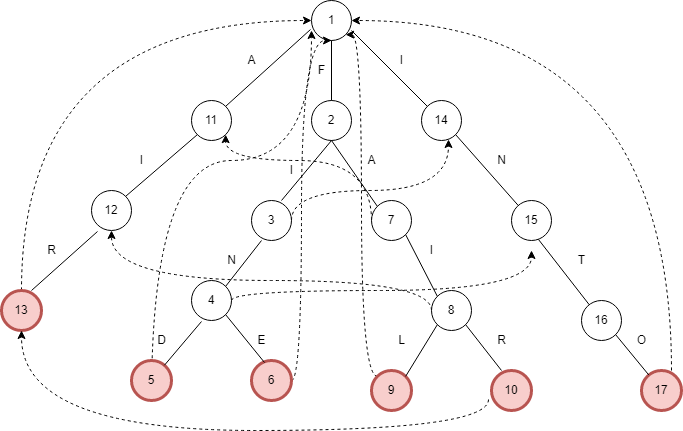}
\caption{The Aho-Corasick multiple pattern matching automaton for the set of keywords \emph{\{AND, FIND, FINE, FAIL, FAIR, INTO\}}. Mismatch transitions are depicted through dashed lines. Leaves are depicted as bolded and colour marked.}
\label{fig:AhoCorasick}
\end{figure}

The deterministic finite automaton (DFA) proposed by Aho-Corasick \cite{relatedwork-ahocorasick} is a very popular approach to solving the multiple pattern-matching problem. The Aho-Corasick automaton is built from a trie tree, constructed from the set of keywords that need to be looked up in an input data set. The trie tree is then transformed into a deterministic finite automaton (DFA) by computing a "mismatch" transition (failure function) that points to the longest suffix of the word at the current node that is also present in the trie. The mismatch transition is used whenever there is no match of the input set at the current node. Every node will save information about the word associated with the respective position inside the trie (prefix), and the transition towards the next state. For example, for the automaton in Figure \ref{fig:AhoCorasick}, node 4 is associated with the word \emph{FIN},  has the suffixes \emph{[IN, N, <empty>]}, and while \emph{IN} is also a word in the trie, pointing to node 15, the mismatch transition of node 4 will therefore point to node 15.

One drawback of the Aho-Corasick automaton is that, while it performs very well in tracking accurate matches of patterns in an input string, it is not suitable to track alterations of those patterns in the input data. Nonetheless, its almost linear runtime speed makes it one of the top choices for exact pattern-matching nowadays, and that was the starting premises for the non-deterministic finite automaton (NDFA) approach that we propose in this paper.

\section{Implementation}
\label{Implementation}

\subsection{Our NDFA-based approach}

Our approach is inspired by the Aho-Corasick \cite{relatedwork-ahocorasick} finite state machine, however it differs significantly from it as it creates a new NDFA state machine, using a sliding window concept at node-level, as well as a locally-applied similarity measurement for the sliding window concept that we introduce. We modify the failure transitions of the NDFA, and we are changing the entire heuristics of those transitions so that we can apply inexact pattern-matching. In order to achieve this, we allow the transitions from a certain state to have multiple outcomes when parsing the same input characters. In order to start determining the output of any given transition, we employed the usage of the sliding window concept, accompanied by the computation of every possible suffix for every node in our automaton.

In order to achieve this, we improve upon the classic Aho-Corasick DFA, which formally can be described as a finite ordered list of 5 symbols $M=\{Q, \sigma, \delta, q_{i}, F\}$, and transform it into a NDFA 5-tuple $M^{'}=\{Q, \sigma, \delta, q_{i}, F\}$. In both approaches, the Q symbol represents the finite set of states, $\sigma$ is representative of the used alphabet, $q_{i}$ is the start state, and F is the set of final states. The main difference between the formally described aforementioned methods is implementation of the transition set $\delta$. In the classical approach, the transitions can occur from a single state to another single state,based on the input characters, and can be formally described as $\delta \colon Q \times \sigma \xrightarrow{} Q $. In our approach, the transitions from a certain state can have multiple outcomes by parsing the same input characters, thus the transition definition will translate to $\delta \colon Q \times \sigma \xrightarrow{} P(Q) $, where P(Q) is a power-set of Q, or a set of possible combinations of states in which the transition will end up. In order to start determining the output of any given transition, we employed the usage of the sliding window concept, accompanied by the computation of every possible suffix for every node in our automaton.

From an implementation point-of-view, our proposed NDFA will add a new pre-processing step, computed after the failure functions calculation inside of the automaton. In this newly added pre-processing stage, for every node in our automaton, we will start parsing every possible transition, or when that is not possible, the failure function linked to that node, so that in the end, we computed all the viable suffixes of a length equal to that of a predetermined sliding window length, able to start from that node. This approach effectively produces, for every node, a list (as long as the sliding window length) of all possible matches from that particular node forward, significantly reducing matching comparisons when the local optimum is employed.

\begin{figure}
\centering
\includegraphics[scale=0.45]{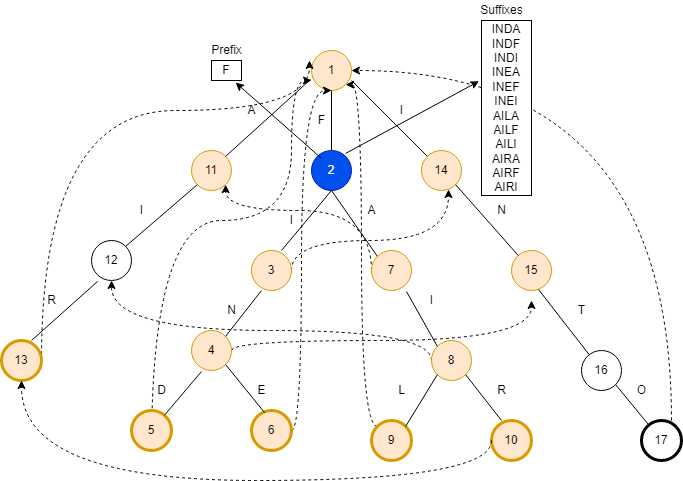}
\caption{The non-deterministic multiple pattern matching automaton for the same set of keywords \emph{\{AND, FIND, FINE, FAIL, FAIR, INTO\}}, using a sliding window of length 4. Parsed nodes used for computing suffixes are colour marked.}
\label{fig:NDFA_AhoCorasick}
\end{figure}

For example, in figure \ref{fig:NDFA_AhoCorasick}, for the node 2, the preprocessing stage of computing the sliding window length suffixes will generate the suffixes list \emph{[INDA, INDF, INDI, INEA, INEF, INEI, AILA, AILF, AILI, IRN]}. In memory-constrained environments, the introduction of this additional storage will prove a challenge of its own, one that opens new opportunities for future work and development.

\begin{figure}
\centering
\includegraphics[scale=0.35]{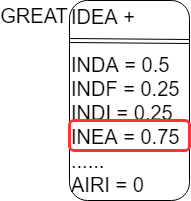}
\caption{Computation of Hamming distance at node 2 for all the possible suffixes}
\label{fig:Hamming}
\end{figure}

\subsection{Local optimum threshold computation}
The nondeterministic behavior is properly highlighted in the actual processing stage. At this stage, the novelty brought by this NDFA approach is the possibility of applying local similarity measurements for approximate pattern-matching at any given step inside the parsing of the automaton for the length of the sliding window, instead of computing them at every position inside the input, for the length of the entire keyword at the processing stage. We chose to focus on the Euclid, Hamming and Levenshtein distances, as the similarity measures used for the local optimum threshold computation process. Figure ~\ref{fig:Hamming} shows the comparisons done at node 2 when the input \emph{GREATIDEA} is parsed, and also the select of the best result from all similarity scores computed. By obtaining a locally determined result of the similarity metric applied at any given node (between a concatenation of the node's prefix and the suffixes, one at a time), and the string of the same length resulting by parsing the input to get to the respective node, we can decide if we move throughout the automaton with an existing transition (similar to the exact match of the Aho-Corasick DFA), or we jump to the mismatch transition, thus allowing multiple outcome states for the same input. The distinction between the aforementioned cases is done by an \textit{a priori} definition of a local optimum threshold value for all the similarity metrics results. We hypothesize that choosing a high value for this local optimum threshold may result in certain keywords not being detected (false negatives), while setting the value too low, might produce other matches than expected (false positives). In order to try to mitigate the occurrence of false positives, we also added an additional verification so no match is detected if more than a given empirical percentage of the length is disjoint. The percentage is set based on the maximum length of the keywords and the sliding windows. For example, for a maximum keyword length of 10 characters, the existing similarity measurements for approximate pattern-matching are using a percentage of 20\%, while for the same keyword length size and a sliding window of 5, in order to have related similarities, the percentage of our approach need to be set as 10\%.

\begin{figure}
\centering
\includegraphics[scale=0.55]{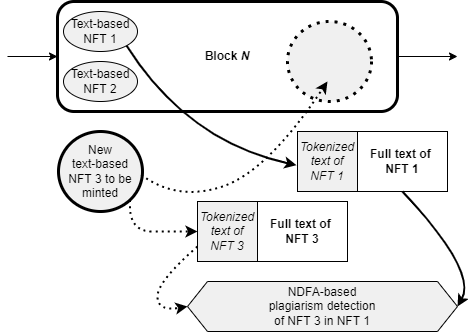}
\caption{Proposed implementation and architectural layout of a text-based NFT in a blockchain ledger.}
\label{fig:implementation-nft}
\end{figure}

\subsection{Blockchain NFT minting}
While NFTs may have become popular due to digital art (in particular, photographs), they may also hold other types of information, including books, various texts, novels, short stories, etc., all of which are meant to ensure that specific form of digital art can be later traded or otherwise preserved in the blockchain. In our proposed approach, in order for an NFT to be minted, its associated digital data needs to be verified against plagiarism. After it passes verification, it can be stored in the ledger as minted. For text-based NFTs, we propose an approach where the original text is stored alongside the tokenized text. This way, tokenized text can be verified using our proposed NDFA-based approach against plagiarism, with other already-minted NFTs. As speed is a concerning factor in all blockchain-drive platforms, our focus on this aspect becomes stronger and more clear. A proposed layout of the text-based NFTs in a blockchain ledger is outlined in Figure ~\ref{fig:implementation-nft}.

\subsection{Challenges and results interpretation}
The ideal situation for this scenario, as opposed to employing the traditional similarity measures in text mining, would be to obtain no false negatives and no false positives. The next best fit here would be the situation where we have false positives, but no false negatives - as the false positives can be trialed in a separate, post-processing step. The worst-case scenario is the situation where false negatives occur in large amounts, offsetting by a large margin the number of results of the classic similarity measures.

\section{Experimental results}
\label{ExperimentalResults}

\begin{figure}
\centering
\begin{subfigure}{.5\textwidth}
  \centering
  \includegraphics[width=1\linewidth]{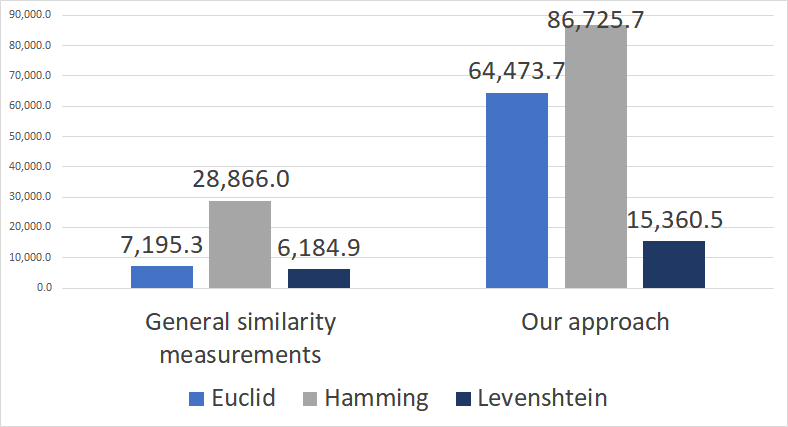}
  \caption{Throughput (Kbps)}
  \label{fig:Throughput1}
\end{subfigure}%
\begin{subfigure}{.5\textwidth}
  \centering
  \includegraphics[width=1\linewidth]{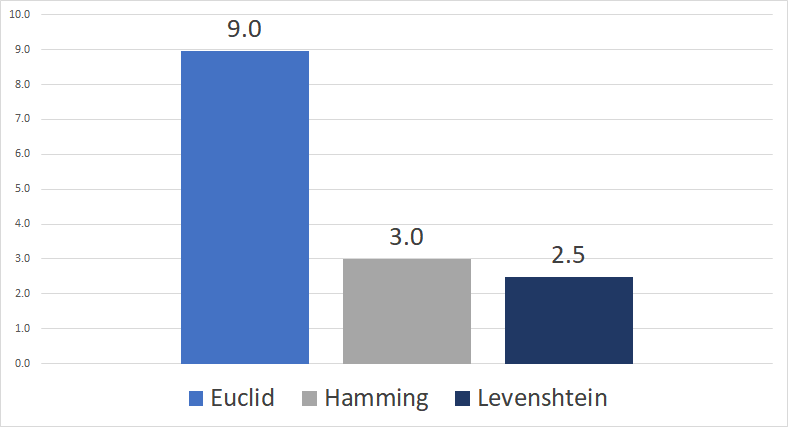}
  \caption{Speed-up ($\times$ times) }
  \label{fig:Speedup1}
\end{subfigure}
\caption{Performance comparison between classic similarity measures and our proposed approach in Scenario 1}
\label{fig:DataSet1}
\end{figure}

We have performed our experiments on an AMD Ryzen 5 5600H CPU, with 16 GB of DDR4 RAM and a 512 GB SSD drive, running on Windows 10. For the experimental dataset, we have used slightly modified versions of  datasets created by Paul Clough (Information Studies) and Mark Stevenson (Computer Science), at the University of Sheffield \cite{experimentalresults-dataset,experimentalresults-dataset-download}. For scenario 1, we benchmarked the performance of our approach using a pair of datasets of 75 patterns (of variable length ranging between 4 to 15 characters), and an input size of 2,450 KB. We also ran consecutive tests in order to empirically determine the best local optimum threshold values, so that the results produced by our approach match the ones produced by the classic similarity measurements used: Euclid, Hamming and Levenshtein. Figure \ref{fig:DataSet1} showcase the improvement in terms of throughput, as well as the speed-up when benchmarking the performances of our approach and the basic similarity measurements for the Euclid, Hamming and Levensthein distances computation.

\begin{figure}
\centering
\includegraphics[scale=0.35]{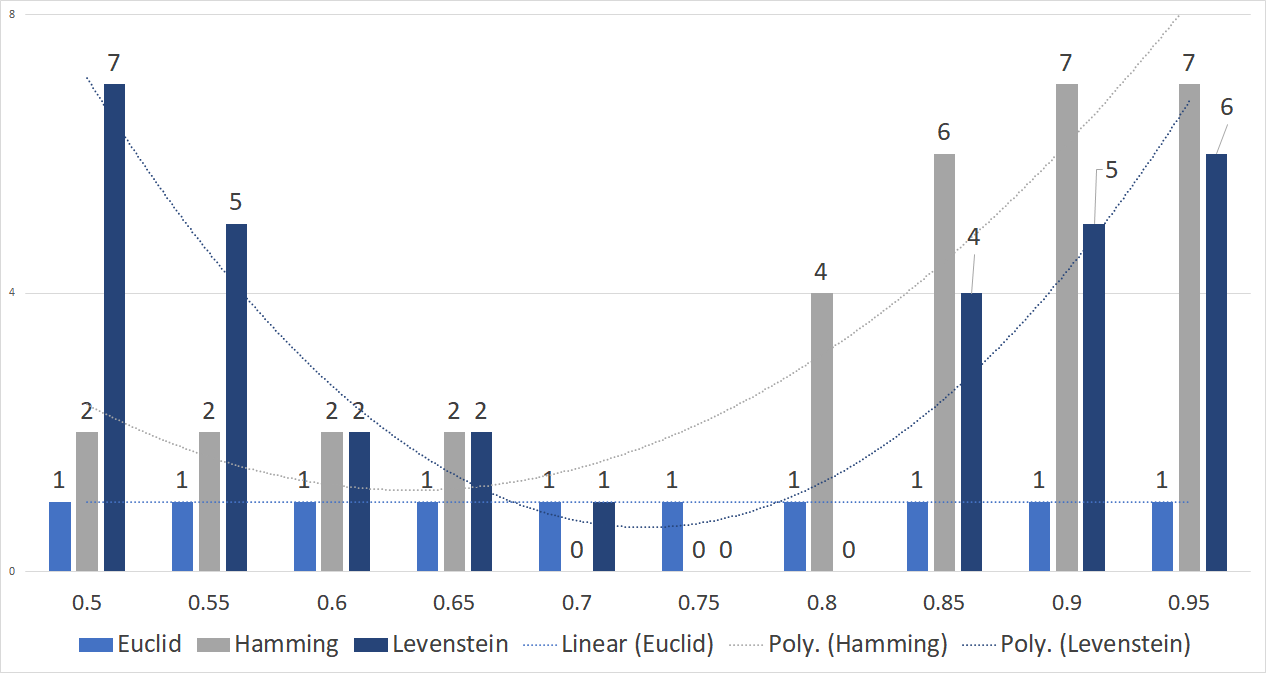}
\caption{Number of false negatives, for various local threshold values (0.5 to 0.95).}
\label{fig:FalseNegatives}
\end{figure}

\begin{figure}
\centering
\includegraphics[scale=0.35]{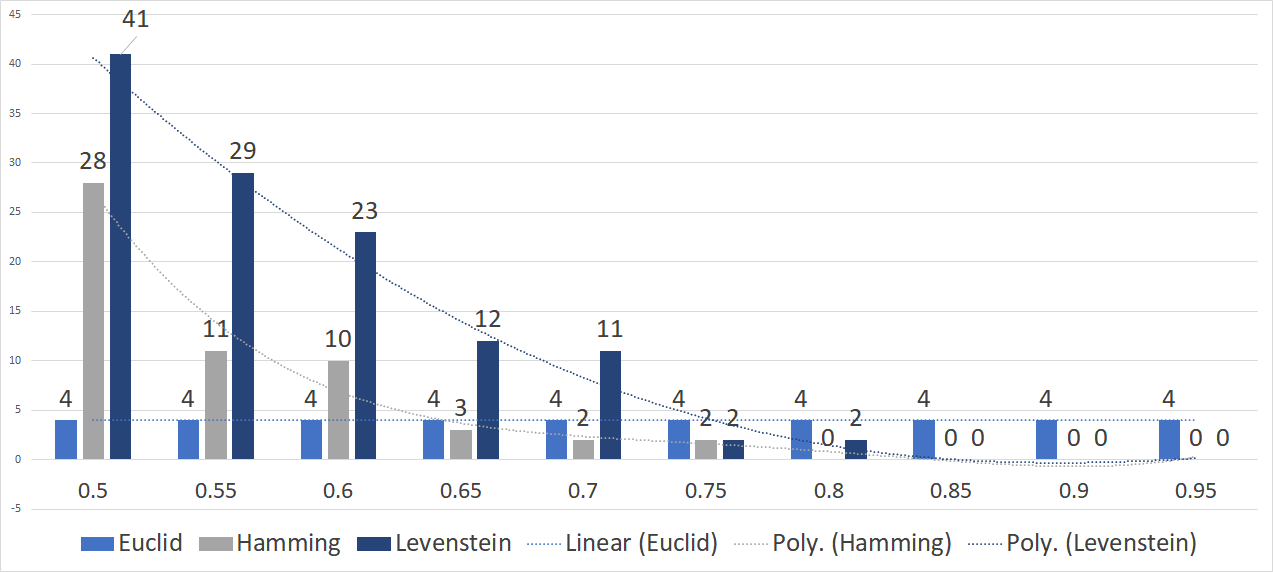}
\caption{Number of false positives, for various local threshold values (0.5 to 0.95).}
\label{fig:FalsePositives}
\end{figure}

Figures \ref{fig:FalseNegatives} and \ref{fig:FalsePositives} showcase how the outcome is influenced by changing the local optimum threshold values. We correctly presumed that setting the value too high will cause false negatives. Also, due to the increased number of false positives, when the local optimum threshold is set too low, we would be dealing with an additional set of false negatives. This is caused by partial matches both in the sliding window and the input data, which produce better local similarity scores as opposed to the global score, associated with the entire pattern. Therefore, we recommend that the sliding window length is at least as high as the length of the longest pattern in our set of keywords.

In terms of accuracy, the number of false negatives in Figure ~\ref{fig:FalseNegatives} is a constant 1 achieved for the Euclidean similarity, throughout the entire experiment. Both Hamming and Levensthein similarities return no false negatives when the threshold value is set to 0.75, which seems to be the best fit for the scenarios tested. The number of false positives remains constant for the Euclidean similarity, but follows a decreasing trend as the threshold values get closer to 1 for the others (which, in theory, would produce the same exact matching results as in the original Aho-Corasick DFA) - particularly, starting with a threshold value of 0.85, both Hamming and Levenshtein similarities produce no false positives.

\begin{figure}
\centering
\begin{subfigure}{.5\textwidth}
  \centering
  \includegraphics[width=1\linewidth]{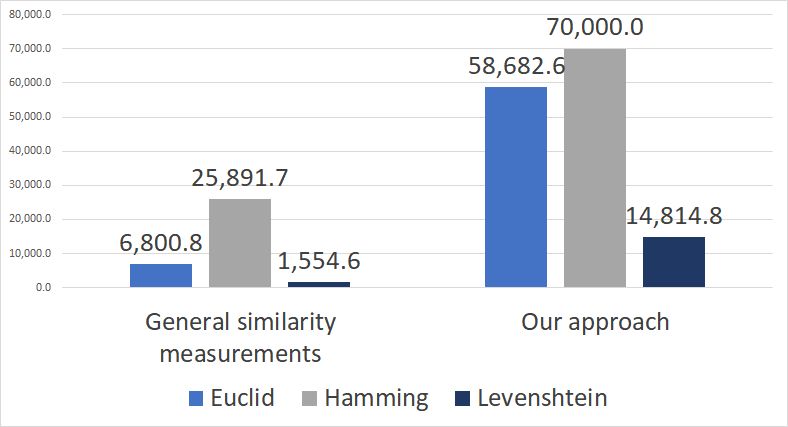}
  \caption{Throughput (Kbps)}
  \label{fig:Throughput2}
\end{subfigure}%
\begin{subfigure}{.5\textwidth}
  \centering
  \includegraphics[width=1\linewidth]{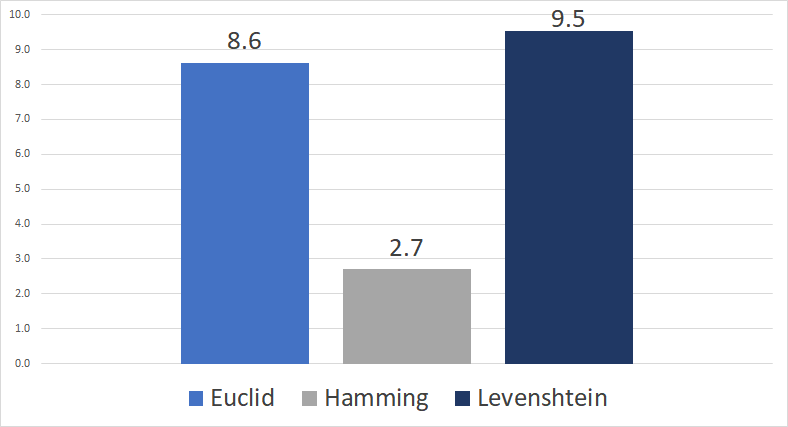}
  \caption{Speed-up ($\times$ times) }
  \label{fig:Speedup2}
\end{subfigure}
\caption{Performance comparison between classic similarity measures and our proposed approach in Scenario 2}
\label{fig:DataSet2}
\end{figure}

For thorough testing, we have empirically observed that the longer the patterns in the keyword set tend to be, the faster the speed-up achieved for the Levensthein similarity. The results of this particular setup (which we call here as scenario 2), with patterns of double lengths as in the original scenario 1, have been outlined in Figure \ref{fig:DataSet2}. We have observed significant run-time speed-ups for our proposed approach, compared to the classic approaches. In particular, for the dataset tested in scenario 2, the Levenshtein similarity measurement throughput has improved to about 9.5$\times$ higher throughput, up from 2.5$\times$ higher throughput as in scenario 1. The Euclid-driven similarity measurement has achieved a speed-up of about 9$\times$ in both scenarios tested, and similarly the Hamming distance provided around 3$\times$ higher throughput in our approach.

One additional challenge that we are currently looking to overcome, is the reduction of the storage size required for our proposed approach. While the throughput performance is significantly better than that the classic approaches, our approach uses significantly more memory due to the size of the NDFA automaton, including the storage of all suffixes at each node level. We are already working on an improved storage model for our approach therefore, which will be the subject of a future paper.

\section{Conclusions}
\label{Conclusions}

In this paper, we have presented an innovative fast approach to performing approximate-pattern matching for plagiarism detection, with particular applicability to blockchain-driven, NFT-ready platforms and ecosystems. For our proposed implementation, we have used our own NDFA-based automaton along with a sliding window concept and local thresholds at node-level, for tracing partial matches faster. We have tested our approach and concluded that it behaves suitably similar to existing various other similarity measures used in text mining for plagiarism detection, while obtaining a significant speed-up improvement of up to 9.5$\times$ faster throughput. This makes our approach significantly faster than classic similarity measurements, and provides a potential approach to integrating the technology with blockchain-driven NFTs, in order to further secure the uniqueness of the non-fungible tokens and reduce the risk of plagiarism.

\subsection{Future work} Future work includes various aspects of intrinsic functionality being tested, including: a) more similarity measurements being used, tested and potentially developed; b) the corroboration of a semantic parser to the approximate pattern-matching performed in the lexical parsing process; c) the reduction of storage requirements for large sliding window sizes, and applicability of our approach to heterogeneous architectures; d) applicability of our methodology to other fields of research and data science.

\subsection{Acknowledgements} This research was supported by the virtuaLedger project \cite{virtualedger} and the MOISE project number 240/2020, ID POC/398/1/1, financed by EU, Romanian government and West University of Timisoara. The views expressed in this paper do not necessarily reflect those of the corresponding project’s partners.


\end{document}